\documentclass[paper]{JHEP3} 

\usepackage{epsfig,multicol,multirow}
\usepackage{amsmath,subfigure,latexsym,amssymb,cite}
\usepackage{here}
\usepackage{bbm}

\newcommand\fverb{\setbox\fverbbox=\hbox\bgroup\verb}
\newcommand\fverbdo{\egroup\medskip\noindent%
			\fbox{\unhbox\fverbbox}\ }
\newcommand\fverbit{\egroup\item[\fbox{\unhbox\fverbbox}]}
\newbox\fverbbox

\newcommand {\beq} {\begin{equation}}
\newcommand {\eeq} {\end{equation}}
\newcommand {\beqa}{\begin{eqnarray}}
\newcommand {\eeqa}{\end{eqnarray}}

\newcommand {\tr}{{\rm tr\,}}

\newcommand{\1}{\mbox{1}\hspace{-0.25em}\mbox{l}}

\allowdisplaybreaks

\title{Late time behaviors of the expanding universe in the IIB matrix model}

\author{Sang-Woo Kim,$^{a}$
Jun Nishimura${}^{bc}$ and Asato Tsuchiya${}^{d}$
\vspace*{0.5cm} \\
\llap{$^a$}Department of Physics, Osaka University,\\
Toyonaka, Osaka 560-0043, Japan\\
\llap{$^b$}Department of Particle and Nuclear Physics,\\
Graduate University for Advanced Studies (SOKENDAI),\\
Tsukuba, Ibaraki 305-0801, Japan\\
\llap{$^c$}High Energy Accelerator Research Organization (KEK),\\
Tsukuba, Ibaraki 305-0801, Japan\\
\llap{$^d$}Department of Physics, Shizuoka University,\\
836 Ohya, Suruga-ku, Shizuoka 422-8529, Japan
\vspace*{0.5cm} \\
\email{sang@het.phys.sci.osaka-u.ac.jp, jnishi@post.kek.jp, satsuch@ipc.shizuoka.ac.jp}}

\preprint{OU-HET-752-2012\\ KEK-TH-1563}

\abstract{
Recently we have studied
the Lorentzian version of the IIB matrix model
as a nonperturbative formulation of
superstring theory.
By Monte Carlo simulation, we have shown that the notion of time
---as well as space---emerges dynamically from this model,
and that we can \emph{uniquely} extract the real-time dynamics,
which turned out to be rather surprising:
after some ``critical time'',
the SO(9) rotational symmetry of the
nine-dimensional
space is
spontaneously broken down to SO(3) and
the three-dimensional space starts to expand rapidly.
In this paper, we study the same model based on
the classical equations of motion,
which are expected to be valid at later times.
After providing a general prescription to solve the equations,
we examine a class of solutions, which correspond
to manifestly commutative space.
In particular,
we find a solution with an expanding behavior
that naturally solves the cosmological constant problem.
}


\begin{document}

\section{Introduction}
\label{sec:introduction}

There are many fundamental questions in cosmology,
which can, in principle, be answered by superstring theory.
Describing the birth of the universe is one of the most fundamental ones.
%
It is recognized, however, that
the cosmic singularity is not resolved generally
in perturbative string theory \cite{Liu:2002ft,Liu:2002kb,Lawrence:2002aj,%
Horowitz:2002mw,Berkooz:2002je}.
Therefore, in order to study the very early universe,
we definitely need a nonperturbative formulation.
Among various proposals \cite{BFSS,IKKT,DVV} based on
matrix models,\footnote{For earlier attempts to apply matrix models
to cosmology, see ref.\ \cite{Freedman:2004xg,Craps:2005wd,Li:2005sz,%
Das:2005vd,Chen:2005mga,She:2005mt,Martinec:2006ak,Ishino:2006nx,%
Matsuo:2008yd,Klammer:2009ku,Lee:2010zf}.}
the IIB matrix model \cite{IKKT} looks most natural
for describing the birth of the universe,
since not only space but also time is
expected to appear
dynamically from the matrix degrees of freedom.
Also the model is unique in that it is a manifestly covariant formulation,
while the other proposals are based on the light-cone formulation,
which breaks covariance.

One of the important issues in the IIB matrix model
is to identify the configurations of matrices that dominate
the path integral
and to determine the corresponding space-time structure.
This was studied by various
approaches \cite{AIKKT,Hotta:1998en,Ambjorn:2000bf,%
Ambjorn:2000dx,Anagnostopoulos:2001yb,Anagnostopoulos:2011cn,%
Nishimura:2000ds,Nishimura:2000wf,Nishimura:2001sx,Kawai:2002jk,Aoyama:2006rk,%
Imai:2003vr,Imai:2003jb,Imai:2003ja,Nishimura:2011xy}
in the Euclidean version of the model,
which was shown to have finite partition function
without any cutoffs \cite{Krauth:1998xh,Austing:2001pk}.
However, the Euclidean model does not seem suitable for cosmology
since it does not provide the real-time dynamics.
Furthermore, a recent study based on the gaussian expansion method
suggests that the space-time obtained dynamically in the Euclidean model
is three-dimensional rather than four-dimensional \cite{Nishimura:2011xy}.
While this conclusion itself may have a profound implication,
we do not know yet how it should be physically interpreted.

All these considerations led
us to study
the Lorentzian version of
the IIB matrix model nonperturbatively \cite{KNT}.
By Monte Carlo simulation, we have shown that
the Lorentzian model can be made well-defined nonperturbatively
by first introducing infrared cutoffs and then removing
them appropriately in the large-$N$ limit.
We have also found that the eigenvalue distribution of
the matrix in the temporal direction
extends
in that limit,
which implies that time emerges dynamically.
(Supersymmetry of the model plays a crucial role.)
Indeed we were able to extract a unique real-time dynamics,
which turns out to have a surprising property.
After some critical time,
the SO(9) rotational symmetry of the space is broken spontaneously
down to SO(3), and the three-dimensional space starts to expand rapidly.
This result can be interpreted as the birth of the universe.
Note that the concept of time-evolution also emerges dynamically
in this model without ever having to specify the initial condition.

Although the length of time-evolution that has been extracted from
Monte Carlo simulation is restricted due to finite $N$,
we consider that the whole history of the universe
can be obtained from
the same model in the large-$N$ limit.
If this is true, we should be able to answer various important questions
in both cosmology and particle physics.
For instance, we may obtain the microscopic description
of inflation.
If we can reach the time at which stringy excitations
and quantum gravitational effects
become negligible, we may see how particles
in the Standard Model
(or possibly its extension) starts to appear.
%
If we can study the behaviors of the model at much later times,
we may be able to understand
why the expansion of our universe is accelerating in the present epoch.
Finally we may even predict how our universe will be like in the future.
%
%
%

While the late-time behaviors are difficult to study by
direct Monte Carlo methods,
the classical equations of motion are expected to become more and more
valid at later times
since the value of the action increases with the cosmic expansion.
%
%
%
%
%
%
%
We will see that there are actually
many classical solutions,
which is reminiscent of the fact
that string theory possesses infinitely many vacua
that are perturbatively stable.
However, unlike in perturbative string theory,
we have the possibility to pick up the unique solution
that describes our universe by requiring
smooth connection
to the behavior at earlier times
accessible by
Monte Carlo simulation.
%
From this perspective,
we consider it important to
classify the
classical solutions and to examine their cosmological implications.
The aim of this paper is
to make a first step in that direction.
In particular, we find a classical solution with an expanding
behavior that can naturally solve the cosmological constant problem.
%
%
%

Another issue we would like to address
in this paper
concerns
how the commutative space-time appears
from this model.
%
%
This is important since
the SO(9) symmetry breaking
observed in the Monte Carlo simulation
is understood intuitively by a mechanism,
which relies crucially on the fact that the matrices that represent
the space-time are noncommutative \cite{KNT}.
%
We show that
classical solutions which correspond to manifestly commutative space
can be easily constructed,
and discuss the vanishing of
noncommutativity between space and time
in some simple examples.
This implies that the emergence of commutative space-time
is indeed possible in this model at later times.

The rest of this paper is organized as follows.
In section \ref{sec:IIBmm}
we briefly review the IIB matrix model and the results
obtained by our previous Monte Carlo studies \cite{KNT}.
In section \ref{sec:classical-sol}
we provide a general prescription
to find classical solutions.
In particular, we show that this can be done systematically
by using Lie algebras.
Then we restrict ourselves to manifestly space-space commutative solutions
and present a complete classification of such solutions
within certain simplifying ansatz.
The particular solution discussed
in our previous publication \cite{KNT2}
also appears in this classification.
In section \ref{sec:su11-su2}
we obtain explicit time-evolution of the scale factor and
the Hubble parameter for some simple solutions,
and discuss their cosmological implications.
Section \ref{sec:concl} is devoted to a summary and discussions.
In appendix \ref{sec:unitary-rep}
we review
the irreducible unitary representations of the SU$(1,1)$ algebra,
which will be needed in constructing the solutions discussed
in section \ref{sec:su11-su2}.
In appendix \ref{sec:others}
we list the other simple solutions which are not discussed
in section \ref{sec:su11-su2}.
In appendix \ref{sec:not-ss-com}
we give some examples of solutions that are not manifestly
space-space commutative.

\section{The IIB matrix model and the birth of the universe}
\label{sec:IIBmm}

In this section we discuss
why the IIB matrix model is considered
as a nonperturbative formulation of type IIB superstring theory
in ten dimensions, and how the results suggesting
the birth of the universe were obtained
by our previous Monte Carlo studies of this model.

The action of the IIB matrix model is given as \cite{IKKT}
%
\beqa
S &=& S_{\rm b} + S_{\rm f} \ ,
\label{action}
 \\
S_{\rm b} &=& -\frac{1}{4g^2} \, \tr \Bigl( [A_{\mu},A_{\nu}]
[A^{\mu},A^{\nu}] \Bigr) \ ,
\label{b-action}
\\
S_{\rm f}  &=& - \frac{1}{2g^2} \,
\tr \Bigl( \Psi _\alpha (\, {\cal C} \,  \Gamma^{\mu})_{\alpha\beta}
[A_{\mu},\Psi _\beta] \Bigr)  \ ,
\label{f-action}
\eeqa
where  $A_\mu$ ($\mu = 0,\cdots, 9$) and
$\Psi_\alpha$ ($\alpha = 1,\cdots , 16$) are
$N \times N$
Hermitian matrices.
The Lorentz indices $\mu$ and $\nu$ are
contracted
using the metric
$\eta={\rm diag}(-1 , 1 , \cdots , 1)$.
The $16 \times 16$ matrices $\Gamma ^\mu$ are
ten-dimensional gamma matrices after the Weyl projection,
and the unitary matrix ${\cal C}$ is the
charge conjugation matrix.
The action has manifest SO(9,1) symmetry,
where $A_{\mu}$ and $\Psi _\alpha$ transform as a
vector and
a Majorana-Weyl spinor, respectively.

There are various evidences
that the model gives a nonperturbative formulation of
superstring theory.
First of all, the action (\ref{action})
can be viewed as a matrix regularization of the worldsheet action
of type IIB superstring theory in a particular gauge known
as the Schild gauge \cite{IKKT}.
It has also been argued that
configurations of block-diagonal matrices
correspond to a collection of disconnected worldsheets
with arbitrary genus.
Therefore, instead of being
equivalent
just to the worldsheet theory,
the large-$N$ limit of the matrix model is expected to be
a second-quantized theory of type IIB superstrings,
which includes multi-string states.
Secondly, D-branes are represented as classical solutions in the
matrix model, and the interaction between them calculated
at one loop reproduced correctly the known results
from type IIB superstring theory \cite{IKKT}.
Thirdly, one can derive
the light-cone string field theory for the type IIB case
from the matrix model \cite{FKKT} with a few assumptions.
In the matrix model, one can define
the Wilson loops, which can be naturally identified
with the creation and annihilation operators of strings.
Then, from the Schwinger-Dyson equations for the Wilson loops,
one can actually obtain the string field Hamiltonian.

In all these connections to string theory,
it is crucial that the model has two kinds of
fermionic symmetries given as
\begin{align}
& \left\{ \begin{array}{ll}
\delta^{(1)}A_{\mu} &=i\epsilon_1 {\cal C}\Gamma_{\mu} \Psi  \ , \\
\delta^{(1)}\Psi &=\frac{i}{2}\Gamma^{\mu\nu}[A_{\mu},A_{\nu}]\epsilon_1 \ ,
\end{array} \right.
\label{susy1} \\
& \left\{ \begin{array}{ll}
\delta^{(2)}A_{\mu} &=0 \ , \\
\delta^{(2)}\Psi &=\epsilon_2 \1 \ ,
\end{array} \right.
\label{susy2}
\end{align}
where $\1$ is the unit matrix.
It also has the bosonic symmetry given by
\begin{align}
\left\{ \begin{array}{ll}
\delta^{(3)}A_{\mu} &=c_{\mu} \1  \ , \\
\delta^{(3)}\Psi &=0 \ .
\end{array} \right.
\label{translation}
\end{align}
Let us denote the generators of (\ref{susy1}),
(\ref{susy2}) and (\ref{translation}) by
$Q^{(1)}$, $Q^{(2)}$ and $P_{\mu}$, respectively,
and define their linear combinations
\begin{align}
\tilde{Q}^{(1)}=Q^{(1)}+Q^{(2)} \ ,
\;\;\;
\tilde{Q}^{(2)}=i(Q^{(1)}-Q^{(2)}) \ .
\end{align}
Then, we find that the generators satisfy the algebra
\begin{align}
[\epsilon_1{\cal C}\tilde{Q}^{(i)},\epsilon_2{\cal C}\tilde{Q}^{(j)}] =
-2\delta^{ij}\epsilon_1{\cal C}\Gamma^{\mu}\epsilon_2P_{\mu} \ ,
\label{N2SUSY}
\end{align}
where $i,\, j=1,2$.
This is nothing but the ten-dimensional ${\cal N}=2$ supersymmetry.
It is known that field theories with
this symmetry
necessarily include gravity, which suggests that
so does the IIB matrix model.
When we identify (\ref{N2SUSY}) with
the ten-dimensional ${\cal N}=2$ supersymmetry,
the symmetry (\ref{translation}) is identified
with the translational symmetry in ten dimensions,
which implies that
the eigenvalues of $A_{\mu}$ should
be identified with
the coordinates of ten-dimensional space-time.
This identification
is consistent with the one adopted in stating
the evidences listed in the previous paragraph,
and shall be used throughout this paper as well.

An interesting feature of the IIB matrix model
is that the space-time itself is treated as a part of
dynamical degrees of freedom in the matrices.
One can therefore try to
identify the dominant configurations of matrices
in the path integral and to determine the corresponding
space-time structure.
For that purpose, one needs to define the partition function
as a finite matrix integral.
This is nontrivial since the bosonic part of the action
(\ref{b-action}) is not bounded from below.
By decomposing it into two terms
\beq
S_{\rm b} = -\frac{1}{2g^2} \, \tr ( F_{0i} )^2 +
\frac{1}{4g^2} \, \tr ( F_{ij} )^2  \ ,
\eeq
where we have defined Hermitian matrices
$F_{\mu\nu} = i [A_\mu , A_\nu]$,
we find that the first term is negative, whereas the
second term is positive.
A common way to overcome this problem is to make a Wick rotation,
which amounts to making replacements
$A_0 \mapsto iA_{10}, \;\;\; \Gamma^0 \mapsto -i\Gamma^{10}$.
The Euclidean version of the model obtained in this way
is well-defined nonperturbatively without
any cutoffs \cite{Krauth:1998xh,Austing:2001pk}.
The space-time structure has been studied by various approaches
in this Euclidean model \cite{AIKKT,Hotta:1998en,Ambjorn:2000bf,%
Ambjorn:2000dx,Anagnostopoulos:2001yb,Anagnostopoulos:2011cn,%
Nishimura:2000ds,Nishimura:2000wf,Nishimura:2001sx,Kawai:2002jk,Aoyama:2006rk,%
Imai:2003vr,Imai:2003jb,Imai:2003ja,Nishimura:2011xy}.
However, it is nontrivial whether the Wick rotation is valid
in a theory including gravity \cite{Ambjorn:2005qt,Kawai:2011rj}.
Furthermore, the recent result based on the gaussian expansion
method \cite{Nishimura:2011xy}
suggests that the space-time appearing dynamically
in the Euclidean model is three-dimensional rather than four-dimensional.

For these reasons, we studied the Lorentzian model
nonperturbatively for the first time in ref.~\cite{KNT}.
In order to make the partition function finite,
we introduced the infrared cutoffs by imposing the following constraints
\begin{align}
&\frac{1}{N}\mbox{tr}(A_0)^2 \leq \kappa \frac{1}{N}\mbox{tr}(A_i)^2 \ ,
\nonumber\\
&\frac{1}{N}\mbox{tr}(A_i)^2 \leq L^2 \ ,
\label{constraints}
\end{align}
where $\kappa$ and $L$ are the cutoff parameters.
We have shown by Monte Carlo simulation that these cutoffs can be removed
in the large-$N$ limit in such a way that the physical quantities scale.
The resulting theory thus obtained has no parameters
other than one scale parameter.
This feature is precisely what one expects
for nonperturbative string theory.

It turned out that not only space but also time emerges dynamically
in this Lorentzian model.
We found that the eigenvalue distribution
of $A_0$ extends in the large-$N$ limit.
Here supersymmetry of the model plays a crucial role.
If we omit fermions, the eigenvalue distribution has a finite extent.
Furthermore, it turned out that one can extract
the real-time dynamics
by working in the SU($N$) basis which diagonalizes $A_0$.
We found that after a critical time,
the SO(9) symmetry is spontaneously broken down to the SO(3),
and three out of nine spatial directions start to expand.
This can be interpreted as the birth of the universe.
Note that the real-time dynamics is an emergent notion in this model,
and we do not even have to specify the initial conditions.
The above result is unique in that sense.

\section{Classical solutions in the Lorentzian model}
\label{sec:classical-sol}

\subsection{General prescription to find classical solutions}
\label{sec:general-prescription}

In this subsection we present
a general prescription to find classical solutions
in the Lorentzian model.
It is important to take into account that the two cutoffs
had to be introduced in order to make the model well-defined
as we reviewed in the previous section.
Since the inequalities (\ref{constraints}) are actually saturated
as is also seen by Monte Carlo simulation \cite{KNT},
we search for stationary points of
the bosonic action $S_{\rm b}$
for fixed $\frac{1}{N} \tr (A_0)^2$ and $\frac{1}{N} \tr (A_i)^2$.
Then we have to extremize the function
\begin{align}
\tilde{S}=\mbox{tr}
\left( - \frac{1}{4}[A_{\mu},A_{\nu}][A^{\mu},A^{\nu}]
+\frac{\tilde{\lambda}}{2}(A_0^2-\kappa L^2)
-\frac{\lambda}{2}(A_i^2-L^2)\right) \ ,
\label{tildeS}
\end{align}
where $\lambda$ and $\tilde{\lambda}$ are the Lagrange multipliers.

Differentiating (\ref{tildeS}) with respect to $A_0$ and $A_i$,
we obtain\footnote{Classical solutions in the
IIB matrix model have been studied
in ref.~\cite{Steinacker:2011wb} for $\lambda=\tilde{\lambda}=0$.
They are studied in the Euclidean version in
ref.~\cite{Chatzistavrakidis:2011su}.}
\begin{align}
-[A_0,[A_0, A_i]]+[A_j,[A_j, A_i]] - \lambda A_i &= 0 \ , \label{cl-eq1} \\
[A_j,[A_j, A_0]]  - \tilde{\lambda} A_0 &= 0 \ ,
\label{cl-eq2}
\end{align}
respectively.
Differentiating (\ref{tildeS})
with respect to $\tilde{\lambda}$ and $\lambda$,
we obtain
\begin{align}
&\frac{1}{N}\mbox{tr}(A_0^2) = \kappa L^2 \ , \label{A0constraint}\\
&\frac{1}{N}\mbox{tr}(A_i^2) = L^2 \ ,  \label{Aiconstraint}
\end{align}
respectively.
Once we obtain a solution to eqs.~(\ref{cl-eq1}) and (\ref{cl-eq2}),
we can substitute them
into (\ref{A0constraint}) and (\ref{Aiconstraint})
to determine $\lambda$ and $\tilde{\lambda}$
as a function of $\kappa$ and $L$.

Here we point out that the terms
in eqs.~(\ref{cl-eq1}) and (\ref{cl-eq2})
proportional to the Lagrange multipliers break
the ${\rm SO}(9,1)$ symmetry in general.
In fact, we will see that there is a set of solutions
with $\lambda = \tilde{\lambda}$,
which do not suffer from this explicit breaking.
However, we do not impose this condition from the outset
in order to keep our analysis as general as possible.

In what follows, we consider solutions with
\begin{align}
A_i=0 \;\;\; \mbox{for} \;\; i >d  \ ,
\label{ansatz}
\end{align}
where $d\leq 9$.
This is motivated from our observation in Monte Carlo simulation
that $A_i$ in the extra dimensions remain small when the three-dimensional
space expands.
From this point of view, one may think that we should choose $d=3$.
However, one can actually construct a solution with larger $d$
by taking a direct sum of solutions with smaller $d$, say with $d=1$,
as we will see in \ref{sec:d1}.
We therefore keep $d$ arbitrary for the moment.

A general prescription to solve
the equations of motion (\ref{cl-eq1}) and (\ref{cl-eq2})
within the ansatz (\ref{ansatz})
is given as follows.
Let us first define a sequence of commutation relations
\begin{align}
[A_i,A_j]&=iC_{ij} \ , \label{C}\\
[A_i,C_{jk}]&=iD_{ijk}\ , \label{D}\\
[A_0,A_i]&=iE_i \ , \label{E}\\
[A_0,E_i]&=iF_i \ , \label{F}\\
[A_i,E_j]&=iG_{ij} \ , \  \cdots  \ , \label{G}
\end{align}
where $1 \leq i,j,k \leq d$ and the symbols
on the right-hand side represent
Hermitian operators newly defined.
Then we determine the relationship
among $A_0$, $A_i$, $C_{ij}$, $D_{ijk}$, $E_i$, $F_i$, $G_{ij},\cdots$
so that
the equations of motion (\ref{cl-eq1}) and (\ref{cl-eq2})
and the Jacobi identities are satisfied.
We obtain a Lie algebra in this way.
Considering that all the operators are Hermitian,
each unitary representation of the Lie algebra
gives a classical solution.

\subsection{Lie algebra for manifestly commutative space}
\label{sec:ss-com}

Since we expect that commutative space appears
at later times,
we restrict ourselves to solutions
corresponding to manifestly commutative space in what follows.
(See appendix \ref{sec:not-ss-com} for
examples of solutions without this restriction.)
This implies that we impose
\begin{align}
[A_i,A_j]=0 \ .
\label{commutative}
\end{align}
Then we follow the general prescription described
in section \ref{sec:general-prescription}.
In particular, we show
that one can actually close the algebra with a finite number
of generators by imposing a simple condition.



Let us consider the relationship among
$A_0$, $A_i$, $C_{ij}$, $D_{ijk}$, $E_i$, $F_i$, $G_{ij}$
in (\ref{C}) $\sim$ (\ref{G}).
First of all, (\ref{commutative}) implies that
\begin{align}
C_{ij}=0 \ , \;\;\; D_{ijk}=0 \ .
\end{align}
It is convenient to make the irreducible decomposition
of a $d$-dimensional second-rank tensor $G_{ij}$
in (\ref{G}) as
\begin{align}
G_{ij}=M_{ij}&+ N_{ij}+ \frac{1}{d} \delta_{ij}H \ , \label{AE} \\
\mbox{where} \quad
M_{ij}&=M_{ji} \ , \;\;\; \sum_{i=1}^dM_{ii}=0 \ , \label{M}\\
N_{ij}&=-N_{ji} \ .
\end{align}
From the equations of motion (\ref{cl-eq1}) and
(\ref{cl-eq2}), we obtain
\begin{align}
F_i &=\lambda A_i \ , \label{F_i} \\
H   &=\tilde{\lambda}A_0 \ ,
\label{H}
\end{align}
where we have used (\ref{E}), (\ref{F}), (\ref{G}), (\ref{commutative})
and (\ref{AE}).

Next we consider the Jacobi identities
\begin{align}
[A_0,[A_i,A_j]]+[A_i,[A_j,A_0]]+[A_j,[A_0,A_i]]&=0 \ ,
\label{jacobi1} \\
[A_0,[A_i,E_j]]+[A_i,[E_j,A_0]]+[E_j,[A_0,A_i]]&=0 \ ,
\label{jacobi2} \\
[A_i,[E_j,E_k]]+[E_j,[E_k,A_i]]+[E_k,[A_i,E_j]]&=0 \ ,
\label{jacobi3} \\
[E_i,[A_j,A_k]]+[A_j,[A_k,E_i]]+[A_k,[E_i,A_j]]&=0 \ .
\label{jacobi4}
\end{align}
Using  (\ref{E}), (\ref{G}), (\ref{commutative}) and (\ref{AE})
in (\ref{jacobi1}), we find
\begin{align}
N_{ij}=0 \ .
\end{align}
Using (\ref{F_i}) and (\ref{H}) in (\ref{jacobi2}), we find
\begin{align}
[A_0,M_{ij}]=0 \ , \quad [E_i,E_j]=0 \ .
\end{align}
Similarly, from (\ref{jacobi3}) and (\ref{jacobi4}), we obtain
\begin{align}
[E_j,M_{ki}]
-i
\frac{\lambda\tilde{\lambda}}{d}\delta_{ki}A_j
&=[E_k,M_{ij}]
-i
\frac{\lambda\tilde{\lambda}}{d}\delta_{ij}A_k \ ,
\label{EM} \\
[A_j,M_{ik}] -i \frac{\tilde{\lambda}}{d}\delta_{ki}E_j
&=[A_k,M_{ij}] -i \frac{\tilde{\lambda}}{d}\delta_{ij}E_k \ ,
\label{AM}
\end{align}
respectively.
We can easily verify that the Jacobi identities
\begin{align}
[A_0,[E_i,E_j]]+[E_i,[E_j,A_0]]+[E_j,[A_0,E_i]]&=0 \ , \\
[A_i,[A_j,A_k]]+[A_j,[A_k,A_i]]+[A_k,[A_i,A_j]]&=0 \ , \\
[E_i,[E_j,E_k]]+[E_j,[E_k,E_i]]+[E_k,[E_i,E_j]]&=0
\end{align}
are trivially satisfied, hence giving no new relations among the operators.

Now that all the Jacobi identities among $A_0$, $A_i$ and $E_i$
are satisfied,
let us move on to the Jacobi identities that include $M_{ij}$.
In general we need to introduce some new operators,
which appear from the commutator of $M_{ij}$ and one of
$A_0$, $A_i$, $E_i$.
%
One way to close the algebra
without introducing new operators
is to impose that $M_{ij}$ is diagonal.
Let us denote the diagonal elements as
$M_i \equiv M_{ii}$, which satisfy
\begin{align}
\sum_{i=1}^d M_i=0
\label{sum of M}
\end{align}
due to the traceless condition (\ref{M}).
In what follows, we will see that $A_0$, $A_i$, $E_i$ and $M_i$
form a Lie algebra.

First, we find that (\ref{EM}) and (\ref{AM}) lead to
\begin{align}
[E_i,M_j]&=i\frac{\lambda\tilde{\lambda}}{d}(1-d\delta_{ij})A_i \ , \\
[A_i,M_j]&=i\frac{\tilde{\lambda}}{d}(1-d\delta_{ij})E_i \ .
\end{align}
Applying these commutation relations to the Jacobi identity
\begin{align}
[M_i,[A_j,E_k]]+[A_j,[E_k,M_i]]+[E_k,[M_i,A_j]]=0 \ ,
\end{align}
we obtain
\begin{align}
[M_i,M_j]=0 \ .
\end{align}
To summarize, the commutation relations
among $A_0$, $A_i$, $E_i$ and $M_i$
are obtained as
\begin{align}
&[A_i,A_j]=0 \ , \;\;\; [A_0,A_i]=iE_i \ , \;\;\;
[A_0,E_i]=i\lambda A_i \ , \nonumber\\
&[E_i,E_j]=0\ , \;\;\; [A_i,E_j]=i\delta_{ij}
\left(\frac{\tilde{\lambda}}{d}A_0+M_i \right) \ , \;\;\;
[A_0,M_i]=0 \ , \nonumber\\
&[A_i,M_j]=i\frac{\tilde{\lambda}}{d}(1-d\delta_{ij})E_i \ , \;\;\;
[E_i,M_j]=i\frac{\lambda\tilde{\lambda}}{d}(1-d\delta_{ij})A_i \ ,
\;\;\; [M_i,M_j]=0 \ ,
\label{commutation relations of classical solution}
\end{align}
where $1\leq i,j \leq d$ with $1\leq d \leq 9$.
It is straightforward to verify that all the
remaining Jacobi identities including $M_i$ are satisfied
due to
the commutation relations (\ref{commutation relations of classical solution}).
Thus we obtain the Lie algebra
(\ref{commutation relations of classical solution}),
which gives a class
of manifestly space-space commutative solutions.

\subsection{Some simplifications of the Lie algebra}

In this subsection we consider some special cases of
(\ref{commutation relations of classical solution}),
which corresponds to simple Lie algebras.

\subsubsection{the case with $M_i=0$ and $\tilde{\lambda}=0$}

First we point out that one can set
$M_i=0$ and $\tilde{\lambda}=0$ consistently
in eq.~(\ref{commutation relations of classical solution})
for arbitrary $d$, which results in the Lie algebra
\begin{align}
&[A_i,A_j]=0 \ , \;\;\; [A_0,A_i]=i E_i \ , \;\;\;
[A_0,E_i]=i\lambda A_i, \nonumber\\
&[E_i,E_j]=0 \ , \;\;\; [A_i,E_j]=0 \ .
\label{simplification}
\end{align}
We can further simplify (\ref{simplification})
by setting $E_i=\pm \sqrt{\lambda}A_i$, which
yields\footnote{The Lie algebra (\ref{solvable})
in the $d=3$ and $d=4$ cases correspond to
$A_{4,5}^{ab}$ and $A_{5,7}^{abc}$ in Table I of ref.~\cite{patera},
respectively.}
\begin{align}
[A_i,A_j]=0 \ , \;\;\; [A_0,A_i]=\pm i\sqrt{\lambda}A_i \ .
\label{solvable}
\end{align}
The classical solutions obtained from
this Lie algebra
were studied in ref.~\cite{KNT2}.

\subsubsection{the $d=1$ case}
\label{sec:d1}

The Lie algebra simplifies considerably in the $d=1$ case.
In this case, eq.~(\ref{sum of M}) implies that $M_1=0$.
Thus eq.~(\ref{commutation relations of classical solution}) reduces to
\begin{align}
[A_0,A_1]=iE \ , \;\;\;
[A_0,E]=i\lambda A_1 \ , \;\;\;
[A_1,E]=i\tilde{\lambda}A_0 \ ,
\label{d=1}
\end{align}
where we define $E \equiv E_1$.
Note that the $\tilde{\lambda}=0$ case of (\ref{d=1}) is identical
to the $d=1$ case of (\ref{simplification}).

As we mentioned
in section \ref{sec:general-prescription},
we can use the $d=1$ solutions (\ref{d=1})
to construct new solutions
representing
a higher-dimensional space-time in the following way.
For that, we note that
the equations of motion (\ref{cl-eq1}) and (\ref{cl-eq2})
have ${\rm SO}(9)$ symmetry.
Rotating the solution (\ref{d=1}) by an ${\rm SO}(9)$ transformation,
we obtain an equivalent solution, which has
the $i$-th spatial matrix
given by $r_i A_1 \ (i=1,\cdots,9)$ with $r_i^2=1$.
Taking a direct sum of these solutions with various values of $r_i$ ,
we obtain a new solution:
\begin{align}
A'_0&=A_0\otimes \1_k \ , \nonumber\\
A'_i&=A_1\otimes \mbox{diag}(r^{(1)}_{i},r^{(2)}_{i},\cdots,r^{(k)}_{i}) \ ,
\label{new solution} \\
& \mbox{where} \quad
r^{(m)}_i{}^2=1\;\;(m=1,\cdots,k) \ ,
\label{condition for r}
\end{align}
and $\1_k$ is the $k\times k$ unit matrix.
In particular,
we can construct an ${\rm SO}(D)$ symmetric solution
by requiring that $r^{(m)}$'s be distributed uniformly on
a unit $S^{D-1}$, where $1 \le D \le 9$.
The $D=4$ case would then be a physically interesting solution
which represent $(3+1)$-dimensional space-time
with $R\times S^3$ geometry.
%

Let us consider the case in which $\lambda\neq 0$ and $\tilde{\lambda}\neq 0$.
In this case,
the Lie algebra (\ref{d=1}) can be identified either with
the ${\rm SU}(1,1)$ algebra (the ${\rm SL}(2,R)$ algebra)
\begin{align}
[T_0, T_1] = i T_2 \ ,
\quad [T_2, T_0] = i T_1 \ ,
\quad  [T_1, T_2] = -i T_0 \ ,
\label{SU(1,1)}
\end{align}
or with the ${\rm SU}(2)$ algebra
\begin{align}
[L_1,L_2]=iL_3 \ , \;\;\; [L_2,L_3]=iL_1 \ , \;\;\; [L_3,L_1]=iL_2 \ ,
\label{SU(2)}
\end{align}
depending on the signs of $\lambda$ and $\tilde{\lambda}$ as
follows.\footnote{These two algebras appear in Table I of
ref.~\cite{patera} as $A_{3,8}$ and $A_{3,9}$, respectively.}

~\\ \noindent
(a) $\lambda>0$ and $\tilde{\lambda}>0$ : ${\rm SU}(1,1)$ algebra
\begin{align}
&A_0=a T_2 \ ,\;\;\; A_1=bT_0 \ ,\;\;\; E=cT_1 \ , \nonumber\\
&\lambda=a^2 \ , \;\;\; \tilde{\lambda}=b^2 \ ,\;\;\; ab=c \ .
\end{align}

~\\ \noindent
(b) $\lambda<0$ and $\tilde{\lambda}<0$ : ${\rm SU}(1,1)$ algebra
\begin{align}
&A_0=a T_0 \ ,\;\;\; A_1=bT_1 \ ,\;\;\; E=cT_2 \ , \nonumber\\
&\lambda=-a^2 \ , \;\;\; \tilde{\lambda}=-b^2 \ ,\;\;\; ab=c \ .
\label{SU(1,1) solution}
\end{align}

~\\ \noindent
(c) $\lambda>0$ and $\tilde{\lambda}<0$ : ${\rm SU}(1,1)$ algebra
\begin{align}
&A_0=a T_2 \ ,\;\;\; A_1=bT_1 \ ,\;\;\; E=cT_0 \ , \nonumber\\
&\lambda=a^2 \ , \;\;\; \tilde{\lambda}=-b^2 \ ,\;\;\; ab=c \ .
\end{align}

~\\ \noindent
(d) $\lambda<0$ and $\tilde{\lambda}>0$ : ${\rm SU}(2)$ algebra
\begin{align}
&A_0=a L_3 \ ,\;\;\; A_1=bL_1 \ ,\;\;\; E=cL_2 \ , \nonumber\\
&\lambda=-a^2 \ , \;\;\; \tilde{\lambda}=b^2 \ ,\;\;\; ab=c \ .
\label{SU(2)solution}
\end{align}

The cases in which $\lambda=0$ or $\tilde{\lambda}=0$
are discussed in appendix \ref{sec:others}.
Thus we find that the solutions with $d=1$ are
classified into the nine cases; namely,
(a)$\sim$(d) in this subsection and (i)$\sim$(v) in appendix \ref{sec:others}.
In section \ref{sec:su11-su2}
we discuss cosmological implications
of the above four cases (a)$\sim$(d).


\subsubsection{the $d=2$ case}

As a more complicated example, we discuss
the $d=2$ case of
(\ref{commutation relations of classical solution}).
Since this case will not be discussed further in this paper,
impatient readers may jump into section \ref{sec:su11-su2}.

Let us note that
eq.~(\ref{sum of M}) gives $M_1=-M_2\equiv M$.
For $\lambda\neq 0$ and $\tilde{\lambda}\neq 0$,
we can rescale $A_0$, $A_i$, $E_i$ and $M$ appropriately
so that eq.~(\ref{commutation relations of classical solution}) can be rewritten
as
\begin{align}
&[A_1,A_2]=[E_1,E_2]=[A_1,E_2]=[A_2,E_1]=[A_0,M]=0 \ , \nonumber\\
&[A_0,A_1]=iE_1 \ , \;\;\;
[A_0,A_2]=iE_2 \ ,\;\;\;
[A_0,E_1]=i\mbox{sign}(\lambda)A_1 \ ,\;\;\;
[A_0,E_2]=i\mbox{sign}(\lambda)A_2 \ , \nonumber\\
&[A_1,E_1]=2i(\mbox{sign}(\tilde{\lambda})A_0+M) \ , \;\;\;
[A_2,E_2]=2i(\mbox{sign}(\tilde{\lambda})A_0-M) \ , \nonumber\\
&[A_1,M]=-i\mbox{sign}(\tilde{\lambda})E_1 \ ,\;\;\;
[A_2,M]=i\mbox{sign}(\tilde{\lambda})E_2 \ , \nonumber\\
&[E_1,M]=-i\mbox{sign}(\lambda\tilde{\lambda})A_1 \ , \;\;\;
[E_2,M]=i\mbox{sign}(\lambda\tilde{\lambda})A_2 \ ,
\label{d=2}
\end{align}
where $\mbox{sign}( \ \cdot \ )$ represents the sign function.
We compare the above algebra with the ${\rm SO}(2,2)$ algebra
and ${\rm SO}(4)$ algebra,
which can be expressed in a unified way as
\begin{align}
[L_{\alpha\beta},L_{\gamma\delta}]=
ig_{\alpha\gamma}L_{\beta\delta}+ig_{\beta\delta}L_{\alpha\gamma}
-ig_{\alpha\delta}L_{\beta\gamma}-ig_{\beta\gamma}L_{\alpha\delta} \ ,
\label{SO(2,2) or SO(4)}
\end{align}
where $\alpha,\;\beta,\;\gamma,\;\delta=1,2, 3, 4$
and $g_{\alpha\beta}$ represents the metric.
By making an identification
\begin{align}
&A_0=L_{23} \ , \nonumber\\
&A_1=L_{12}+L_{34} \ , \;\;\;
A_2=L_{12}-L_{34} \ , \nonumber\\
&E_1=g_{22}L_{13}-g_{33}L_{24} \ , \;\;\;
E_2=g_{22}L_{13}+g_{33}L_{24} \ , \nonumber\\
&M=-\mbox{sign}(\lambda)L_{14} \ ,
\label{SO(2,2) and SO(4) solutions}
\end{align}
it is easy to check that eq.~(\ref{d=2}) is satisfied if
\begin{align}
(g_{\alpha\beta})= \left\{ \begin{array}{ll}
\mbox{diag}(1,1,-1,-1) & \ \mbox{for} \ \lambda>0,\;\tilde{\lambda}>0  \ ,\\
\mbox{diag}(1,-1,-1,1) & \ \mbox{for} \ \lambda<0,\;\tilde{\lambda}<0  \ , \\
\mbox{diag}(1,-1,1,-1) & \ \mbox{for} \ \lambda>0,\;\tilde{\lambda}<0 \ , \\
\mbox{diag}(1,1,1,1)   & \ \mbox{for} \ \lambda<0,\;\tilde{\lambda}>0 \ .
\end{array} \right.
\end{align}
The first three cases correspond to the ${\rm SO}(2,2)$ algebra,
whereas the last one corresponds to the ${\rm SO}(4)$ algebra.
A unitary representation of these algebras corresponds to
a classical solution.
More in-depth studies of these solutions
are left for future investigations.

\section{Cosmological implications of some simple classical solutions}
\label{sec:su11-su2}

In this section we discuss the cosmological implications
of the ${\rm SU}(1,1)$ solutions
(a)$\sim$(c) and the ${\rm SU}(2)$ solution (d)
discussed in section \ref{sec:d1}.
As a warming up, we start with the ${\rm SU}(2)$ solution, which
is simpler,
and then move on to the ${\rm SU}(1,1)$ solutions,
which exhibit physically more interesting behaviors.


\subsection{Solutions based on the ${\rm SU}(2)$ algebra --- a warm-up}
\label{sec:SU2-warmup}

Let us consider the solution (\ref{SU(2)solution})
based on the ${\rm SU}(2)$ algebra (\ref{SU(2)}).
As we mentioned in section \ref{sec:general-prescription},
we use the unitary representations.
The irreducible unitary representations of
the ${\rm SU}(2)$ algebra is specified by their spins $J$, which are
non-negative integers or half-integers.
In the spin $J$ representation,
the matrix elements of the generators in eq.~(\ref{SU(2)})
are given by
\begin{align}
(L_1)_{mn}&=\frac{1}{2}\sqrt{(J-n)(J+n+1)}
\delta_{m,n+1}+\frac{1}{2}\sqrt{(J+n)(J-n+1)}\delta_{m,n-1} \ , \nonumber\\
(L_2)_{mn}&=\frac{1}{2i}\sqrt{(J-n)(J+n+1)}\delta_{m,n+1}
-\frac{1}{2i}\sqrt{(J+n)(J-n+1)}\delta_{m,n-1} \ , \nonumber\\
(L_3)_{mn}&=n\delta_{mn} \ ,
\label{matrix elements for SU(2)}
\end{align}
where $-J \leq m,n \leq J$.
From eq.~(\ref{SU(2)solution}),
we find
that $A_0$ is diagonal, whereas $A_1$ has a tri-diagonal structure.
This motivates us to extract the time evolution of space
from the $3 \times 3$ submatrices of $A_0$ and $A_1$ defined as\cite{KNT}
\begin{align}
\bar{A}_0(n)&=a\left( \begin{array}{ccc}
n-1 & 0 & 0 \\
0   & n & 0 \\
0   & 0 & n+1
\end{array} \right)   \ , \\
\bar{A}_1(n)&=\frac{b}{2}\left( \begin{array}{ccc}
0 & \sqrt{(J+n)(J-n+1)} & 0 \\
\sqrt{(J+n)(J-n+1)}   & 0 &
\sqrt{(J-n)(J+n+1)} \\
0   & \sqrt{(J-n)(J+n+1)} & 0
\end{array} \right),
\end{align}
where $-J+1 \leq n \leq J-1$.

Similarly, we consider an ${\rm SO}(4)$ symmetric solution
(\ref{new solution}), where $r^{(m)}_i{}^2=1 \;\;(m=1,\cdots,k)$
and $r^{(m)}$ are uniformly distributed on a unit ${\rm S}^3$.
%
In this case, we define
\begin{align}
\bar{A}'_0(n)&=\bar{A}_0(n)\otimes 1_k \ , \nonumber\\
\bar{A}'_i(n)&=\bar{A}_1(n) \otimes
\mbox{diag}(r^{(1)}_i,\cdots,r^{(k)}_i) \ ,
\label{blocks for SO(4) symmetric solution}
\end{align}
where $\bar{A}'_i(n)$ represents
the structure of space at a discrete time $n$.
We also define the extent of space at a discrete time $n$ by
\begin{align}
R(n) \equiv \sqrt{\frac{1}{3k}\mbox{tr}(\bar{A}'_1(n))^2}
=\sqrt{\frac{b^2}{3}(J(J+1)-n^2)} \ .
\label{R^2 for SU(2)}
\end{align}
%

Let us then discuss the continuum limit, in which we send
$ a \rightarrow 0$ and $J\rightarrow \infty$,
and define the continuum time by $t=na$.
We see from eq.~(\ref{R^2 for SU(2)}) that
a nontrivial dependence of $R$ on $t$ is obtained by
keeping $t_{\rm max}=Ja$ and $\frac{b}{a}=\alpha$ fixed.
This leads to
\begin{align}
 R(t)=R_{\rm max}\sqrt{1-\left(\frac{t}{t_{\rm max}}\right)^2} \ ,
\quad \mbox{where} \ R_{\rm max}=\frac{\alpha t_{\rm max}}{\sqrt{3}} \ .
\label{Rt-su2}
\end{align}

From the fact that $A_0$ is diagonal and $A_1$ has a tri-diagonal structure,
we consider that the space-time noncommutativity vanishes
in the continuum limit.
Let us also note that
the extents of space and time defined by
(\ref{A0constraint}) and (\ref{Aiconstraint}) are given by
$L \sim \alpha t_{\rm max}$ and $\kappa \sim 1/\alpha$.



\begin{figure}[htb]
\begin{center}
\includegraphics[height=6cm]{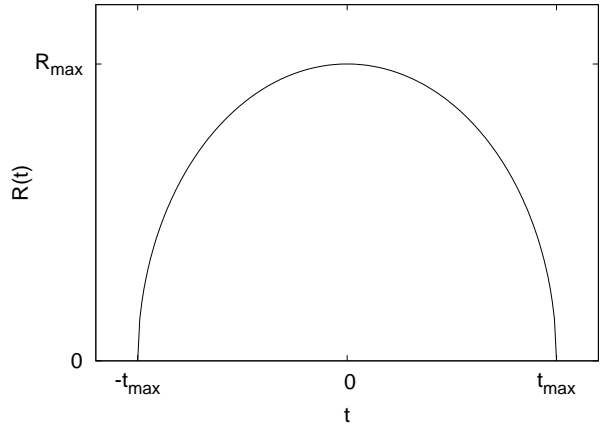}
 \end{center}
\caption{The time dependence of $R(t)$ in the ${\rm SU}(2)$ solution}
\label{su2_rt}
\end{figure}

Let us next discuss the cosmological implication of this solution.
For that we naively identify $R(t)$ with the scale factor
in the Friedman-Robertson-Walker metric.
Then the space-time is $R\times S^3$, where the radius of $S^3$
is given by $R(t)$.
Figure \ref{su2_rt}
shows the time dependence of $R(t)$.
The universe expands towards $t=0$ and shrinks after $t=0$.
The Hubble parameter
can be defined
in terms of the scale factor $R(t)$ as
\begin{align}
H(t)=\frac{\dot{R}(t)}{R(t)}=c \, R(t)^{-\frac{3}{2}(1+w)} \ ,
\end{align}
where $c$ is a constant. From this, we can evaluate the parameter $w$ using
\begin{align}
w=-\frac{2 R(t)}{3}\frac{d\log H(t)}{dR(t)}-1 \ .
\end{align}
Let us recall that $w=1/3$, $w=0$ and $w=-1$ correspond to
the radiation dominated universe,
the matter dominated universe and
the cosmological constant term, respectively.

In the present case of SU(2) solution, we find from (\ref{Rt-su2})
that
\begin{align}
H &=\frac{R_{\rm max}}{t_{\rm max}R^2}\sqrt{R_{\rm max}^2-R^2} \ , \\
w &=\frac{2t_{\rm max}^2}{3t^2}-\frac{1}{3}
\end{align}
for $t<0$.
The parameter $w$ is $w=1/3$ at $t=-t_{\rm max}$,
and it diverges as one approaches $t=0$.


\subsection{Solutions based on the ${\rm SU}(1,1)$ algebra}
\label{sec:su11sol}

Next we discuss the solutions (a), (b) and (c) in section \ref{sec:d1},
which are based on
the ${\rm SU}(1,1)$ algebra (\ref{SU(1,1)}).
We construct ${\rm SO}(4)$ symmetric solutions
for these cases as in eq.~(\ref{new solution}).
The irreducible unitary representations of the ${\rm SU}(1,1)$
algebra are summarized in appendix \ref{sec:unitary-rep}.
Apart from the trivial representation, there are three types:
the primary unitary series
representation (PUSR) (\ref{primary unitary series}),
the discrete series
representation (DSR) (\ref{discrete series 1}) and
(\ref{discrete series 2}),
the complementary unitary series
representation (CUSR) (\ref{complementary unitary series}).
In all these representations, the generator $T_0$ is diagonal.
Hence, we can analytically treat the case (b),
in which $A_0$ is proportional to $T_0$,
while we need to diagonalize $A_0$ numerically in the cases (a) and (c).
For this reason, we start with the case (b).
In this case, we find from (\ref{matrix elements of SU(1,1)}) that
$A_1$ has a tri-diagonal structure.
Therefore, we extract $3\times 3$ submatrices
$\bar{A}_0(n)$ and $\bar{A}_1(n)$ similarly to the ${\rm SU}(2)$ case
discussed in section \ref{sec:SU2-warmup}.
In what follows, we discuss the solutions for each representation separately.

Let us first discuss the solutions corresponding to PUSR.
$\bar{A}_0(n)$ and $\bar{A}_1$ takes the form
\begin{align}
\bar{A}_0(n) &=a\left( \begin{array}{ccc}
n-1+\epsilon & 0 & 0 \\
0   & n+\epsilon & 0 \\
0   & 0 & n+1+\epsilon
\end{array} \right)  \ ,
\label{A0-PUSR}
\\
\bar{A}_1(n)&=\frac{ib}{2}\left( \begin{array}{ccc}
0 & n+i\rho-\frac{1}{2}+\epsilon & 0 \\
-n+i\rho+\frac{1}{2}-\epsilon  & 0 & n+i\rho+\frac{1}{2}+\epsilon \\
0   & -n+i\rho-\frac{1}{2}-\epsilon  & 0
\end{array} \right) \ ,
\end{align}
where $\epsilon=0 \;\mbox{or} \; \frac{1}{2}$
and $\rho$ is a non-negative number, which specifies a representation.

When we consider an SO(4) symmetric solution,
we define $\bar{A}'_0(n)$ and $\bar{A}'_i(n)$
as in eq.~(\ref{blocks for SO(4) symmetric solution}).
Then we find that the extent of space $R(n)$ at a discrete time $n$
becomes
\begin{align}
R(n)=\sqrt{\frac{b^2}{3}\left(n^2+\rho^2+\frac{1}{4}\right)} \ .
\end{align}

Let us take the continuum limit.
We define the continuum time by $t=na$ and take the limit in which
$a\rightarrow 0$ with $\frac{b}{a}=\alpha$ fixed.
We can tune $\rho$ so that $t_0\equiv \rho a$ is fixed.
Then $R(t)$ is given by
\begin{align}
R(t)=\sqrt{\frac{\alpha^2}{3}(t^2 + t_0^2)} \ .
\label{RtPUSR}
\end{align}

As in the ${\rm SU}(2)$ case,
the continuum limit of this solution
represents
a commutative $(3+1)$-dimensional space-time.
In order to evaluate $L$ and $\kappa$ in
(\ref{A0constraint}) and (\ref{Aiconstraint}),
we introduce a cutoff $N$ for the dimension of the representation.
Then, it is easy to see that (\ref{A0constraint})
and (\ref{Aiconstraint}) give
$L\sim Na$ and $\kappa \sim 1/\alpha$.
The infrared cutoff $L$ is removed by sending $N$ to infinity faster
than $1/a$.
The other parameter $\kappa$, which corresponds
to the ratio of $\mbox{tr}(A_0)^2/N$ to $\mbox{tr}(A_i)^2/N$,
is finite
in the continuum limit,
which looks
different from the situation encountered in Monte Carlo studies,
where we had to send $\kappa$ to infinity \cite{KNT}.
This is not so surprising, however,
given that we are looking at different time
regions, and the speed of expansion changes qualitatively depending
on which region we are looking at.

Here we naively identify $R(t)$ with the scale factor again.
Then the space-time
is $R\times S^3$, where the radius $R(t)$ of $S^3$ is time dependent.
From eq.~(\ref{RtPUSR}),
we obtain the Hubble parameter $H$ and the parameter $w$ as
\begin{align}
H&=\frac{\alpha}{\sqrt{3}R^2}\sqrt{R^2 - \frac{\alpha^2t_{0}^2}{3}}  \ , \\
w&= - \frac{2t_{0}^2}{3t^2}-\frac{1}{3} \ ,
\end{align}
We find that $w$ converges to $-\frac{1}{3}$
as $t\rightarrow\infty$, which corresponds to
the expansion of universe with a constant velocity.

\begin{figure}[htb]
\begin{center}
\includegraphics[height=5cm]{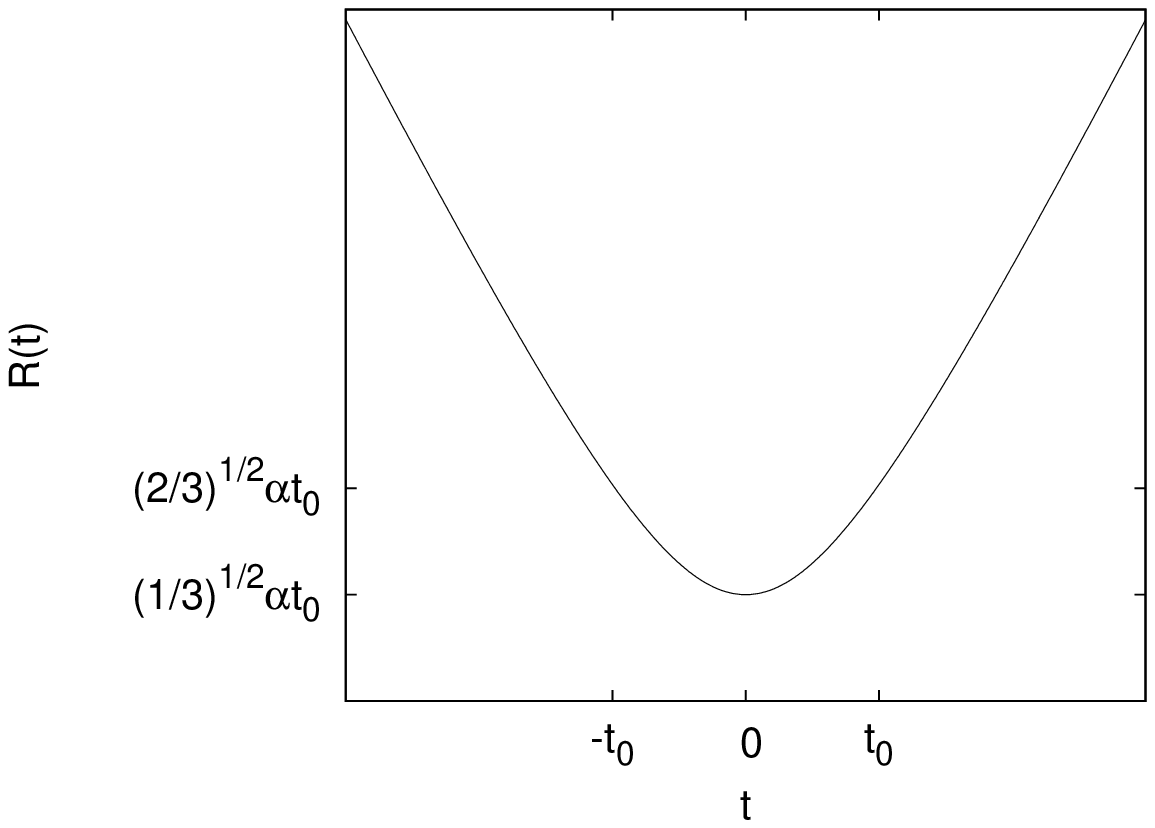}
\includegraphics[height=5cm]{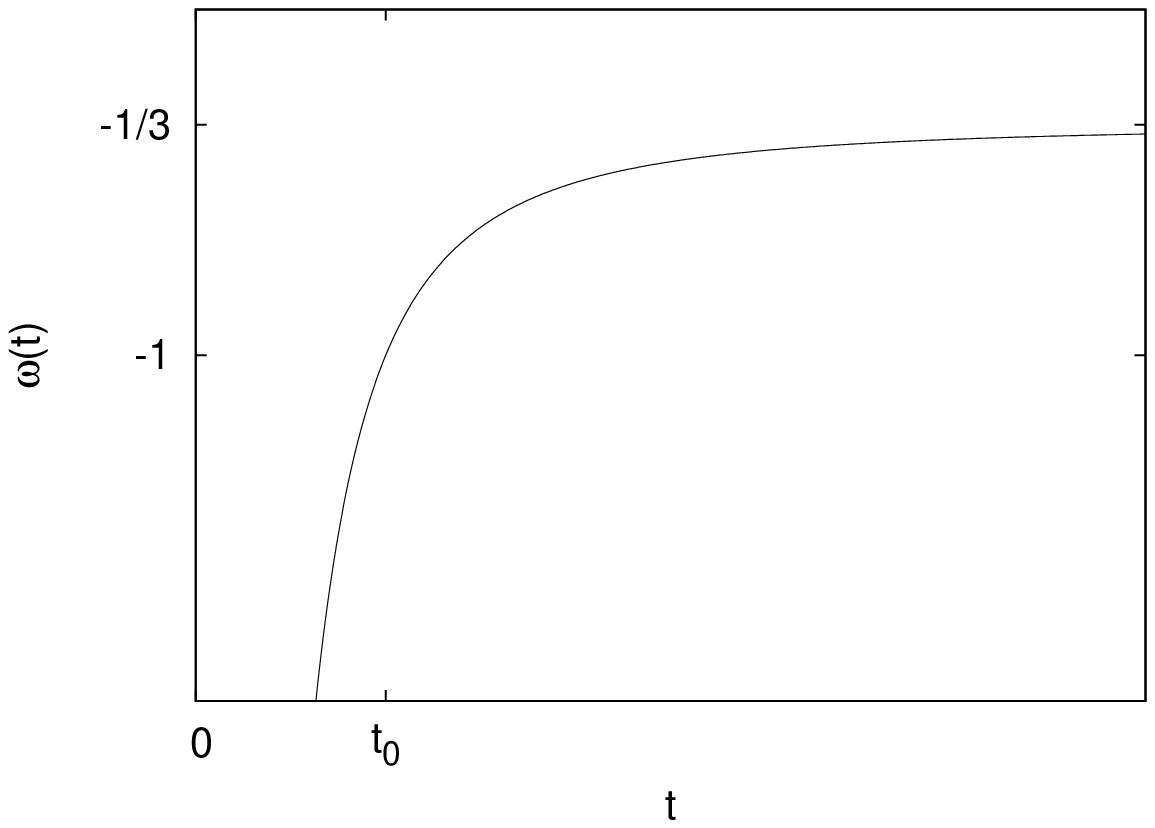}
 \end{center}
\caption{The time dependence of the scale factor $R(t)$ (Left)
and the parameter $w$ (Right)
in the ${\rm SU}(1,1)$ solution with the PUSR.
}
\label{fig:su11pri_rt-wt}
\end{figure}

If we identify $t_0$ with
the present time, the present value of $w$ is $-1$.
This value of $w$ corresponds to
the cosmological constant,
which explains the present accelerating expansion of the universe.
Moreover, the corresponding cosmological constant
becomes of the order of $(1/t_0)^4$,
which naturally solves the cosmological constant problem.
As we mentioned above, $w$ increases with time and approaches $-\frac{1}{3}$.
This means that the cosmological constant actually vanishes in the future.
%
In fig.~\ref{fig:su11pri_rt-wt}
we show the time dependence of $R(t)$ and the parameter $w$.


\vspace{1cm}

Let us next discuss the solutions corresponding to DSR.
$\bar{A}_0(n)$ takes the form (\ref{A0-PUSR}),
where $\epsilon=0 \;\mbox{or} \; \frac{1}{2}$,
and $\bar{A}_1$ takes the form
\begin{align}
\bar{A}_1(n)=\frac{ib}{2}\left( \begin{array}{ccc}
0 & \sqrt{(n+\tau)(n-\tau-1)} & 0 \\
-\sqrt{(n+\tau)(n-\tau-1)}  & 0 & \sqrt{(n-\tau)(n+\tau+1)} \\
0   & -\sqrt{(n-\tau)(n+\tau+1)}  & 0
\end{array} \right) \ ,
\label{A1-CUSR}
\end{align}
with $\tau=-1,-2,-3,\cdots$ for $\epsilon=0$
and $\tau=-\frac{1}{2},-\frac{3}{2},-\frac{5}{2},\cdots$ for
$\epsilon=\frac{1}{2}$.
There is a constraint $n \leq -\tau-\epsilon+1$ or $n \leq \tau-\epsilon-1$.
Then $R(n)$ is given by
\begin{align}
R(n)=\sqrt{\frac{b^2}{3}(n^2-\tau(\tau+1))} \ .
\label{Rn-CUSR}
\end{align}
In the continuum limit,
we can tune $\tau$ so that $t_0\equiv a \tau(\tau+1)$ is fixed.
Then $R(t)$ is given by
\begin{align}
R(t)=\sqrt{\frac{\alpha^2}{3}(t^2 - t_0^2)} \ ,
\label{Rt-DSR}
\end{align}
where the range of $t$ is restricted to either
$t \geq t_0$ or $t \leq -t_0$.

\begin{figure}[htb]
\begin{center}
\includegraphics[height=5cm]{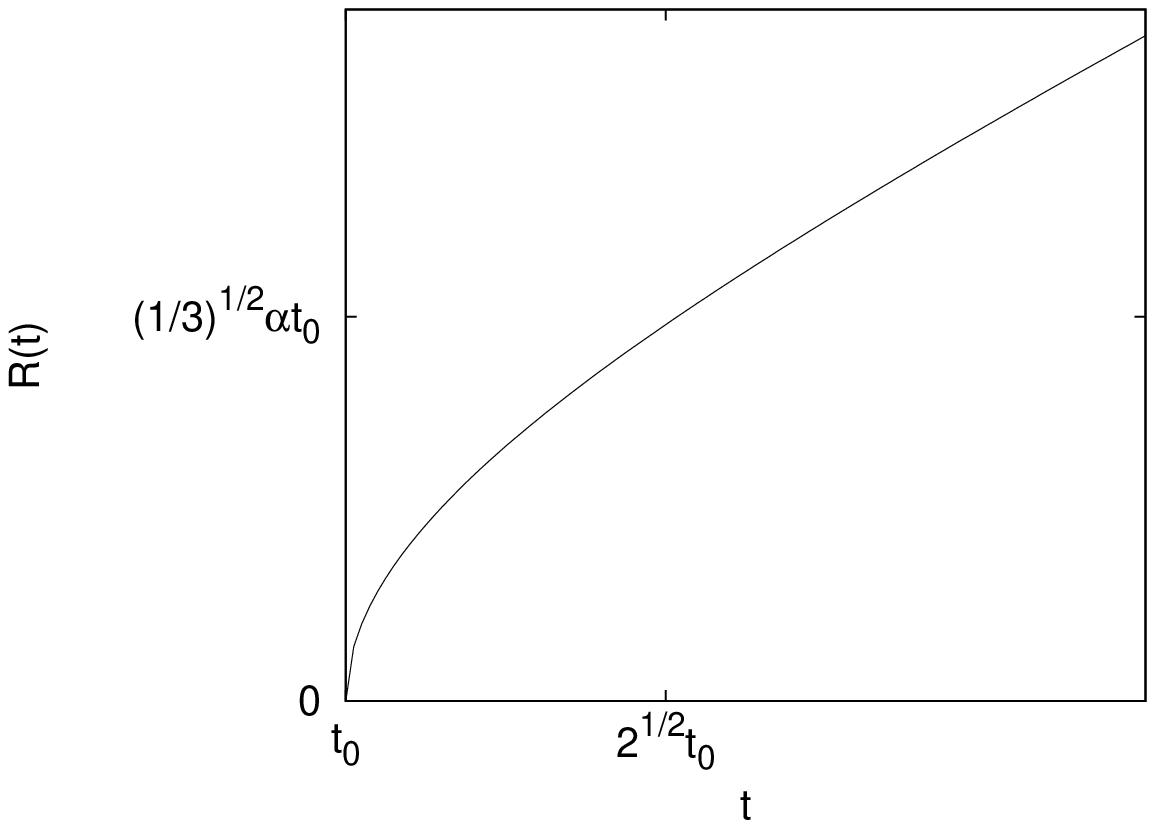}
\includegraphics[height=5cm]{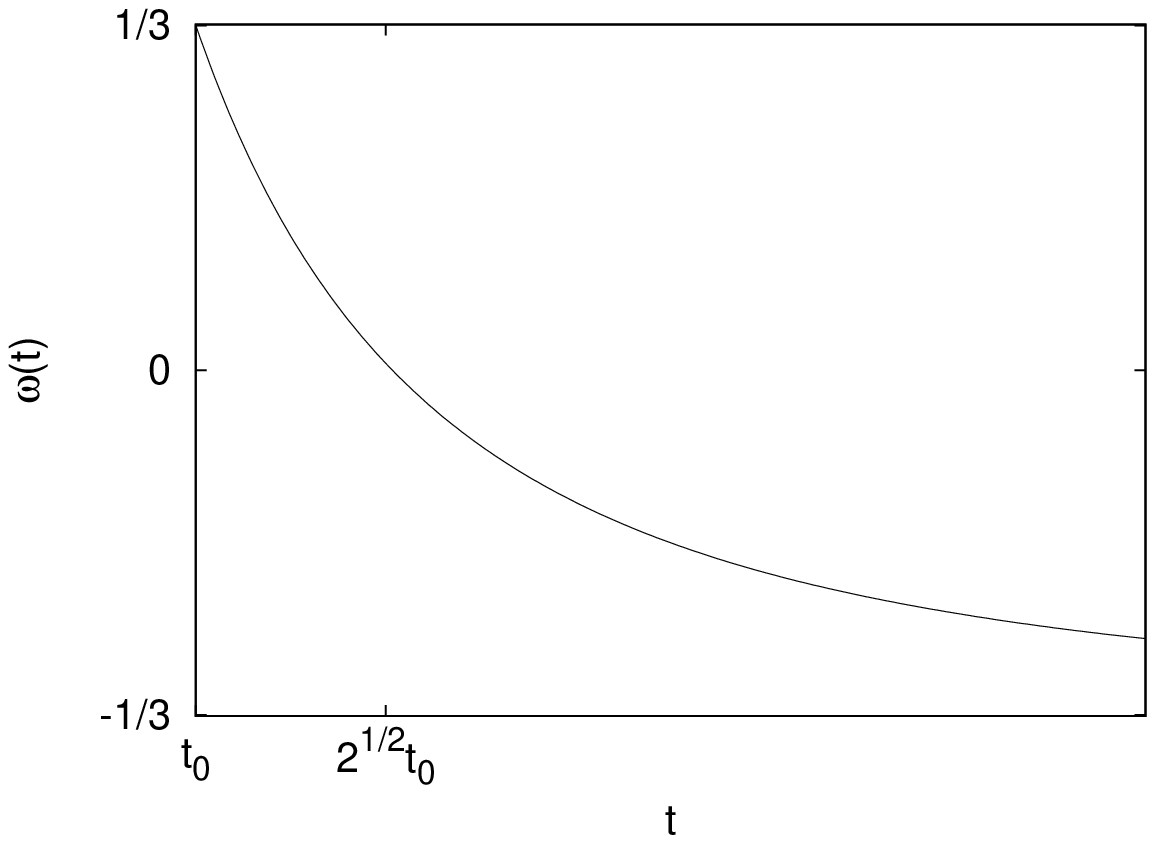}
 \end{center}
\caption{The time dependence of the scale factor $R(t)$ (Left)
and the parameter $w$ (Right)
in the ${\rm SU}(1,1)$ solution with the DSR.
}
\label{fig:su11dis_rt-wt}
\end{figure}

From eq.~(\ref{Rt-DSR}),
we obtain the Hubble parameter $H$ and the parameter $w$ as
\begin{align}
H&=\frac{\alpha}{\sqrt{3}R^2}\sqrt{R^2 + \frac{\alpha^2t_{0}^2}{3}}  \ , \\
w&= \frac{2t_{0}^2}{3t^2}-\frac{1}{3}
\end{align}
for $t \geq t_0$.
We find that the parameter $w$
becomes $w=1/3$ at $t=t_0$ and $w=0$ at $t=\sqrt{2}t_0$.
The former corresponds to the radiation dominant universe,
while the latter to the
matter dominant universe.
Thus this solution may represent some
part of the history of the universe.
Figure \ref{fig:su11dis_rt-wt} shows the time dependence of $R(t)$
and the parameter $w$, respectively.


\vspace{1cm}

Next we discuss the solutions corresponding to CUSR.
$\bar{A}_0(n)$ takes the form (\ref{A0-PUSR}),
where $\epsilon=0$, and $\bar{A}_1(n)$ takes the form (\ref{A1-CUSR}),
where $\tau$ within $-1<\tau <0$ specifies a representation.
The extent of space $R(n)$ is given by eq.~(\ref{Rn-CUSR}).
In the continuum limit, we obtain (\ref{Rt-DSR}), where $t_0=0$.


Finally, we discuss the cases (a) and (c)
in section \ref{sec:d1}.
In these cases, we have diagonalized $A_0$ numerically.
The physical consequence obtained from (a) is essentially
the same as (b).
On the other hand, no accelerating expansion is obtained for the case (c).
$R(t)$ has two peaks, one in the $t<0$ region and the other in
the $t>0$ region, and the minimum lies at $t=0$.

\section{Summary and discussions}
\label{sec:concl}

In this paper we studied the late time behaviors of the universe
in the Lorentzian version of the IIB matrix model.
We investigated the classical equations of motion,
which are expected to be valid at later times.
This is a complementary approach to Monte Carlo simulation,\footnote{See
ref.~\cite{Nishimura:2012xs} for a recent review on Monte Carlo studies
of matrix models and supersymmetric gauge theories in the context of
string theory.}
which was used previously to study the birth of the universe
in the same model.
First we provided a general prescription
to solve the equations of motion.
The problem reduces to that of finding a unitary representation
of a Lie algebra.

In this way, we obtained a class of solutions
that are manifestly space-space commutative.
The simplest ones in this class
are the $d=1$ solutions with $A_2=\cdots=A_9=0$,
from which we can easily construct the ones
representing higher dimensional space-time as well.
We made a complete classification of such solutions.
Some solutions represent expanding $(3+1)$-dimensional universe
without space-time noncommutativity in the continuum limit.
In particular, we find that there exists a solution,
in which the parameter $w$ changes smoothly from $-1$ to $-1/3$.
This explains why we seem to have a tiny cosmological constant
in the present epoch, and
hence can naturally solve the cosmological constant problem.
While we do not insist that this particular solution
really describes our universe, we consider that
the cosmological constant problem can be naturally solved
in the Lorentzian matrix model in a similar manner.


Corresponding to what we have done in Monte Carlo simulation,
we have introduced infrared cutoffs
in both the temporal and spatial directions.
These are represented by the Lagrange multipliers
$\lambda$ and $\tilde{\lambda}$
introduced in the action (\ref{tildeS}).
In general, this breaks the SO(9,1) symmetry of the model explicitly.
Let us note, however, that it is possible to have $\lambda = \tilde{\lambda}$
in the cases (a) and (b) in section \ref{sec:d1}.
Such solutions break the SO(9,1) symmetry \emph{spontaneously}.
We expect that the explicit breaking of the Lorentz symmetry
by the infrared cutoffs disappears in the large-$N$ limit.
If that is really the case, we should select a solution
with $\lambda = \tilde{\lambda}$.
It is intriguing to note that the cases (a) and (b) are indeed the ones
that are physically interesting.

The (3+1)-dimensional space-time represented by the solutions
discussed mainly in this paper has the topology $R\times S^3$.
This is a restriction which we have as long as we construct such
solutions based on the $d=1$ solution.
In other constructions, we can also obtain solutions representing a space
with the topology of a three-dimensional ball as we discussed in
appendix \ref{sec:others}.
While the Monte Carlo results seem to be more consistent with
the latter topology of the space, it remains to be seen
what kind of topology is actually realized at later times.

Below we list some directions for future investigations.

First we consider it important to examine the stability
of the solutions we found in this paper.
It would be also interesting to calculate the one-loop effective action
around the solutions.
That would tell us the validity of the solutions, and we should be
able to know how late the time should be for the solutions to be valid.


Secondly it is important to understand better how one should
extract the information of the space-time metric from a matrix configuration.
Ref.~\cite{Hanada:2005vr} shows that this is indeed
possible, in principle, if one interprets the matrix as a covariant
derivative on the space-time manifold, where
the general coordinate invariance is realized manifestly
as a subgroup of the SU($N$) symmetry.
However, this interpretation is different from the one adopted in this paper,
which is compatible with the supersymmetry
as we reviewed in section \ref{sec:IIBmm}.
The precise relationship between the two interpretations
is yet to be clarified,
although it is tempting to consider that
they
are related to each other by T-duality of type IIB superstring theory.
In this work
we have naively identified $R(t)$
with the scale factor in the Friedman-Robertson-Walker metric
when we discuss cosmological implications in section \ref{sec:su11-su2}.
It remains to be seen whether this identification
can somehow be justified.


Thirdly we consider it important
to study a wider class of solutions
using the general prescription provided in this paper.
In particular, it would be interesting to examine
the solutions, which are not manifestly space-space commutative,
based on the Lie algebra (\ref{SO(2,2) and SO(4) solutions})
or the one given in appendix \ref{sec:not-ss-com}.
Also it would be interesting to investigate solutions
with nontrivial structure in the extra dimensions.
Such structure is expected to play
a crucial role \cite{Chatzistavrakidis:2011gs,Aoki:2010gv}
in determining the matter content at late times
and in finding how the standard model appears from the matrix model.
Eventually, we have to single out the solution, which is smoothly
connected to the unique result at earlier times
accessible by Monte Carlo simulation.

Developments in the above directions would enable us to
solve various fundamental problems in particle physics and cosmology.
For instance, we should be able to understand
the mechanism of inflation and to clarify
what the dark matter and the dark energy are.
We hope that the present work will trigger such developments.

\acknowledgments

We would like to
thank A.~Chatzistavrakidis and T.~Okubo
for valuable discussions.
The authors are also grateful to the participants
of the KITP program ``Novel Numerical Methods for
Strongly Coupled Quantum Field Theory and Quantum Gravity''
for discussions during the meeting.
This research was supported in part by the National Science Foundation
under Grant No.\ NSF PHY05-51164.
S.-W.K.\ is supported by Grant-in-Aid
for Scientific Research
from the Ministry of
Education, Culture, Sports, Science and Technology in Japan (No.\ 20105002).
J.N.\ and A.T.\ is supported in part by Grant-in-Aid
for Scientific Research
(No.\ 19340066, 24540264, and 23244057)
from JSPS.


\appendix

\section{Unitary representations of the ${\rm SU}(1,1)$ algebra}
\label{sec:unitary-rep}
In this appendix we summarize the unitary representations
of the ${\rm SU}(1,1)$ algebra (\ref{SU(1,1)})
based on ref.~\cite{Vilenkin-Klimyk}.
The generators are realized in the space of square integrable functions
in the region $[0,2\pi]$, which we denote as ${\cal L}^2(0,2\pi)$ in
what follows. They are given as
\begin{align}
{\cal T}_0&=i\frac{d}{d\theta}+\epsilon \ ,\nonumber\\
{\cal T}_1&=\frac{i}{2}\left[(\tau+\epsilon)e^{i\theta}
+(\tau-\epsilon)e^{-i\theta}-2\sin\theta\frac{d}{d\theta}\right] \ ,
\nonumber\\
{\cal T}_2&=\frac{1}{2}\left[-(\tau+\epsilon)e^{i\theta}+
(\tau-\epsilon)e^{-i\theta}-2i\cos\theta\frac{d}{d\theta}\right] \ ,
\end{align}
where $0\leq\theta <2\pi$, and
$\tau\in\mathbb{C}$ and $\epsilon\in\mathbb{R}$ are parameters.
It is easy to verify that these operators satisfy
the ${\rm SU}(1,1)$ algebra
\begin{align}
[{\cal T}_0,{\cal T}_1]=i{\cal T}_2 \ , \;\;\;
[{\cal T}_2,{\cal T}_0]=i{\cal T}_1 \ ,\;\;\;
[{\cal T}_1,{\cal T}_2]=-i{\cal T}_0 \ .
\end{align}
Taking the set of functions $\{e^{-im\theta};m\in\mathbb{Z}\}$
as a basis of ${\cal L}^2(0,2\pi)$, one obtains
the matrix elements of the generators as
\begin{align}
({\cal T}_0)_{mn}&=(\epsilon +n)\delta_{mn}\ , \nonumber\\
({\cal T}_1)_{mn}&=-\frac{i}{2}(n-\tau+\epsilon)\delta_{m,n+1}
+\frac{i}{2}(n+\tau+\epsilon)\delta_{m,n-1} \ , \nonumber\\
({\cal T}_2)_{mn}&=-\frac{1}{2}(n-\tau+\epsilon)\delta_{m,n+1}
-\frac{1}{2}(n+\tau+\epsilon)\delta_{m,n-1} \ ,
\label{matrix elements of SU(1,1)}
\end{align}
where $m,n \in \mathbb{Z}$.

The unitary irreducible representations are classified as follows.
We denote the matrix elements of $T_{\mu}$ in (\ref{SU(1,1)})
by $(T_{\mu})_{mn}$, which differs from
$({\cal T}_{\mu})_{mn}$, in general, due to some factors introduced
to define the scalar product that realizes the unitarity.
Since ${\rm SU}(1,1)$ is a noncompact group, all the nontrivial
unitary representations are infinite dimensional.

\noindent
1) primary unitary series representations
\begin{align}
&\tau=i\rho-\frac{1}{2}  \;\;\;(\rho \in \mathbb{R}_{\geq 0}) \ ,
\;\;\; \epsilon=0 \;\;\mbox{or}\;\; \frac{1}{2} \ ,
\nonumber\\
&(T_{\mu})_{mn}=({\cal T}_{\mu})_{mn} \ .
\label{primary unitary series}
\end{align}

\noindent
2) complementary unitary series representations
\begin{align}
&-1 < \tau <1 \ , \;\;\; \epsilon=0 \ , \nonumber\\
&(T_{\mu})_{mn}=\left(\frac{\Gamma(\tau-m+1)\Gamma(-\tau-n)}
{\Gamma(\tau-n+1)\Gamma(-\tau-m)}
\right)^{\frac{1}{2}} ({\cal T}_{\mu})_{mn} \ .
\label{complementary unitary series}
\end{align}

\noindent
3) discrete series representations (I)
\begin{align}
&\tau=-1,-2,-3,\cdots \ , \;\;\;\epsilon=0 \ , \nonumber\\
&\mbox{or} \nonumber\\
&\tau=-\frac{1}{2} \ , -\frac{3}{2},-\frac{5}{2},\cdots \ , \;\;\;
\epsilon=\frac{1}{2} \ , \nonumber\\
&(T_{\mu})_{mn}=\left(\frac{\Gamma(\tau+m+\epsilon+1)
\Gamma(-\tau+n+\epsilon)}
{\Gamma(\tau+n+\epsilon+1)
\Gamma(-\tau+m+\epsilon)}\right)^{\frac{1}{2}}
({\cal T}_{\mu})_{mn} \ ,  \nonumber\\
&m,n \geq -\tau-\epsilon \ .
\label{discrete series 1}
\end{align}

\noindent
4) discrete series representations (II)
\begin{align}
&\tau=-1,-2,-3,\cdots, \;\;\;\epsilon=0 \ , \nonumber\\
&\mbox{or} \nonumber\\
&\tau=-\frac{1}{2},-\frac{3}{2},-\frac{5}{2},\cdots \ , \;\;\;
\epsilon=\frac{1}{2} \ , \nonumber\\
&(T_{\mu})_{mn}=\left(\frac{\Gamma(\tau-m-\epsilon+1)
\Gamma(-\tau-n-\epsilon)}
{\Gamma(\tau-n-\epsilon+1)
\Gamma(-\tau-m-\epsilon)}\right)^{\frac{1}{2}} ({\cal T}_{\mu})_{mn} \ ,
\nonumber\\
&m,n \leq \tau-\epsilon \ .
\label{discrete series 2}
\end{align}

\noindent
5) trivial representation
\begin{align}
T_{\mu}=0 \ .
\end{align}

\section{The other classical solutions in the $d=1$ case}
\label{sec:others}

In this appendix
we discuss the solutions
based on the Lie algebra (\ref{d=1}) obtained for $d=1$
when $\lambda=0$ or $\tilde{\lambda}=0$.
The solutions can be
classified into the following five cases.\footnote{The five cases
(i)$\sim$(v) correspond
to $A_{3,1}$, $A_{3,4}$, $A_{3,6}$, $A_{3,6}$ and $A_{3,4}$,
respectively, in Table I of ref.~\cite{patera}.
In particular, $A_{3,1}$ is a nilpotent Lie algebra.}
%

An interesting feature of the $\tilde{\lambda}=0$ case is that
we can construct solutions representing
higher dimensional space-time with topology other than ${\rm S}^{D-1}$.
The reason is that we can rescale $A_1$ and $E$ in (\ref{d=1})
without changing $\lambda$ when $\tilde{\lambda}=0$.
Therefore, we do not need to impose the condition
(\ref{condition for r}) in constructing
the new solution (\ref{new solution}).
For instance, we can distribute $r^{(m)}_i$ uniformly in a three-dimensional
ball $B^3$ so that the solution represents a $(3+1)$-dimensional space-time
with ${\rm SO}(3)$ symmetry.

~\\ \noindent (i) $\lambda=0$ and  $\tilde \lambda=0$ \\
\noindent
The nontrivial irreducible representations are
parametrized by $a \in \mathbb{R} \;\;(a\neq 0)$.
$A_0$ and $A_1$ are given by the operators acting
on the space of functions of $x$ with $L^2$ integrability,
which is denoted by ${\cal L}^2(\mathbb{R})$ in what follows.
The operators are given explicitly as
\begin{align}
A_0=-ia\sqrt{\lambda}\frac{d}{dx} \ , \;\;\; A_1=x \ , \;\;\; E=-a \ .
\end{align}
This solution represents an infinitely long static D-string.

~\\
\noindent (ii) $\lambda > 0$ and $\tilde{\lambda}=0$ \\
\noindent
In the nontrivial irreducible representations,
$A_0$, $A_1$ and $E$ are operators acting on
${\cal L}^2(\mathbb{R})$, which are given by
\begin{align}
A_0=-i\sqrt{\lambda}\frac{d}{dx}  \ , \;\;\;
A_1=a\cosh x + b \sinh x \ , \;\;\;
E=-\sqrt{\lambda}(a\sinh x +b \cosh x) \ ,
\end{align}
where $a,b\in \mathbb{R}$.

For $a=b$, the solution reduces to
$A_1=a \exp(x)$ and $E=\sqrt{\lambda}A_1$,
which corresponds to the $d=1$ case of (\ref{solvable}).
We can construct an ${\rm SO}(3)$ symmetric solution
from this case by the aforementioned procedure.
The solution thus obtained is equivalent to the one
we constructed from the Lie algebra (\ref{solvable}) with $d=3$
in ref.~\cite{KNT2}, which represents
a $(3+1)$-dimensional expanding universe.
For $a\neq b$, we can also construct
an ${\rm SO}(3)$ symmetric solution in the same way.
We have checked numerically that the resulting solution
exhibits essentially the same behavior as the one for $a=b$.

~\\ \noindent (iii) $\lambda < 0$ and $\tilde{\lambda}=0$ \\
\noindent
In the nontrivial irreducible representations,
$A_0$, $A_1$ and $E$ are operators acting on
${\cal L}^2(0,2\pi)$, which are given as
\begin{align}
A_0 = -i \sqrt{-\lambda}\frac{d}{dx} \ , \;\;\;
A_1 = a \cos x + b \sin x \ ,\;\;\;
E=\sqrt{-\lambda}(a\sin x -b\cos x) \ .
\label{cossin}
\end{align}
This case is analytically tractable.
For the ${\rm SO}(3)$ symmetric solution obtained from the Lie algebra
(\ref{cossin}),
we can define $R(t)$ as in the case of
the ${\rm SU}(2)$ and ${\rm SU}(1,1)$ solutions.
We find that $R(t)={\rm constant}$.

~\\ \noindent (iv) $\lambda=0$ and $\tilde \lambda > 0$ \\
\noindent
In the nontrivial irreducible representations,
$A_0$, $A_1$ and $E$ are operators acting on
${\cal L}^2(0,2\pi)$, which are given as
\begin{align}
A_0 = a \cos x + b \sin x \ , \;\;\;
A_1 = -i \sqrt{\tilde \lambda}\frac{d}{dx} \ ,\;\;\;
E=\sqrt{\tilde{\lambda}}(-a\sin x + b \cos x) \ .
\label{cossin2}
\end{align}
We have seen numerically that
the ${\rm SO}(3)$ symmetric solution obtained from (\ref{cossin2})
exhibits almost the same behavior as
the one in (iii).

~\\ \noindent (v) $\lambda=0$ and $\tilde \lambda < 0$ \\
\noindent
In the nontrivial irreducible representations,
$A_0$, $A_1$ and $E$ are operators acting on
$L^2(\mathbb{R})$ :
\begin{align}
A_0 = a \cosh x + b \sinh x \ ,\;\;\;
A_1 = -i \sqrt{-\tilde \lambda}\frac{d}{dx} \ ,\;\;\;
E=\sqrt{-\tilde{\lambda}}(a \sinh x + b \cosh x) \ .
\label{coshsinh2}
\end{align}
We have seen numerically that $R(t)$ in the ${\rm SO}(3)$ symmetric solution
obtained from (\ref{coshsinh2})
exhibits a behavior different from (ii).
The solutions for $b=0$ has an expanding regime only, whereas
the solutions for $a=0$
have both expanding and contracting regimes.

\section{Examples of classical solutions describing noncommutative space}
\label{sec:not-ss-com}
In this appendix we present some examples of classical solutions
which are not manifestly space-space
commutative.
These solutions are based on the Lie algebras
${\rm SO}(6)$, ${\rm SO}(5,1)$ and ${\rm SO}(4,2)$,
and we interpret them as describing $(3+1)$-dimensional universes
with ${\rm SO}(4)$ symmetry.
These Lie algebras obey the commutation relations
\begin{align}
[L_{\alpha\beta},L_{\gamma\delta}]
=ig_{\alpha\gamma}L_{\beta\delta}+ig_{\beta\delta}L_{\alpha\gamma}
-ig_{\alpha\delta}L_{\beta\gamma}-ig_{\beta\gamma}L_{\alpha\delta} \ ,
\end{align}
where $\alpha,\beta,\gamma,\delta=1,2,\cdots,6$.
The non-vanishing components of $g_{\alpha\beta}$ are
\begin{align}
g_{ii}&=1 \;\;\;  (i=1,2,3,4), \nonumber\\
g_{55}&=\left\{ \begin{array}{lll}
              1 & \mbox{for} & {\rm SO}(6), \;{\rm SO}(5,1)  \ , \\
              -1 & \mbox{for} & {\rm SO}(4,2) \ ,
              \end{array}  \right.\nonumber\\
g_{66}&=\left\{ \begin{array}{lll}
              1 & \mbox{for} & {\rm SO}(6)  \ , \\
              -1 & \mbox{for} & {\rm SO}(5,1), \;{\rm SO}(4,2) \ .
              \end{array} \right.
\end{align}
We set
\begin{align}
&A_0=a L_{56} \ , \nonumber\\
&A_i=b L_{5i}\;\;\;(i=1,2,3,4)  \ , \nonumber\\
&A_5,\cdots,A_9=0 \ .
\end{align}
Then it is easy to verify
that the equations (\ref{cl-eq1}) and (\ref{cl-eq2})
are satisfied if
\begin{align}
\lambda&=-a^2g_{55}g_{66}+4b^2g_{55} \ , \nonumber\\
\tilde{\lambda}&=4b^2g_{55} \ .
\end{align}

\end{document}